\documentclass[aps,graphicx,10pt,showkeys,showpacs,twocolumn]{revtex4}
\usepackage{amsmath}
\usepackage{amscd}
\usepackage{graphicx}
\usepackage{subfigure}
\usepackage{appendix}
\usepackage{multirow}
\usepackage{mathrsfs}

\begin{document}

\title[Short Title]{Efficient and controlled symmetric and asymmetric Bell-state transfers in a dissipative Jaynes-Cummings model}

\author{Qi-Cheng Wu$^{1,\footnote{These authors contributed equally to this work.},\footnote{E-mail: wuqi.cheng@163.com}}$}
\author{Yu-Liang Fang$^{1,*}$}
\author{Yan-Hui Zhou$^{1}$}
\author{Jun-Long Zhao$^{1}$}
\author{Yi-Hao Kang$^{2}$}
\author{Qi-Ping Su$^{2}$}
\author{Chui-Ping Yang$^{2,}$\footnote{E-mail: yangcp@hznu.edu.cn}}

\affiliation{$^{1}$Quantum Information Research Center and Jiangxi Province Key Laboratory of Applied Optical Technology, Shangrao Normal University, Shangrao 334001, China\\
$^{2}$School of Physics, Hangzhou Normal University, Hangzhou}

\begin{abstract}
Realizing efficient and controlled state transfer is necessary for
implementing a wide range of classical and quantum information
protocols. Recent studies have demonstrated that both asymmetric
and symmetric state transfer can be achieved by encircling an
exceptional point (EP) in non-Hermitian (NH) systems. However, the
application of this phenomenon has been restricted to scenarios
where an EP exists in single-qubit systems and is associated with
a specific type of dissipation. In this work, we demonstrate
efficient and controlled symmetric and asymmetric Bell-state
transfers by modulating system parameters within a Jaynes-Cummings
model while accounting for atomic spontaneous emission and cavity
decay. The effective suppression of nonadiabatic transitions
enables a symmetric exchange of Bell states irrespective of the
encircling direction. Furthermore, we report a counterintuitive
finding: the presence of an EP is not indispensable for
implementing asymmetric state transfers in NH systems. We achieve
perfect asymmetric Bell-state transfers even in the absence of an
EP, by dynamically orbiting around an approximate EP. Our work
presents an approach to effectively and reliably manipulate
entangled states with both symmetric and asymmetric
characteristics, through the dissipation engineering in NH
systems.
\end{abstract}


\maketitle

\section{Introduction}

Since the discovery of exceptional points
(EPs)~\cite{EPs1,EPs2,EPs3,EPs4}, a plethora of counterintuitive
phenomena have emerged in non-Hermitian (NH)
systems~\cite{NH1,NH2,NH3,experiment-realize2}, encompassing
heightened sensitivity to perturbations~\cite{sensitivity},
loss-induced transparency effects~\cite{lossinduced,
invisibility}, and the existence of distinctive topological
structures~\cite{topology1,topology-encircling}. Recent studies
have also revealed that symmetric (adiabatic) state
transfer~\cite{topology-encircling,Arkhipov-encircling2,wu-encircling}
and  asymmetric (chiral) state
transfer~\cite{bell-encircling,Berry-encircling,Xu-encircling,Li-encircling,Arkhipov-encircling1,Ergoktas-encircling,Tang-encircling,
Hassan-encircling,Feilhauer-encircling} can be accomplished by
adjusting system parameters along a path encircling an exceptional
point (EP). In general, the symmetric transfer between the
eigenstates of a system implies that the state transfer is solely
determined by the initial state and remains unaffected by the
encircling direction. Conversely, the asymmetric transfer between
the system eigenstates indicates that the state transfer is solely
determined by the encircling direction. The discoveries and
applications of symmetric and asymmetric state transfers in
time-modulated NH systems have undoubtedly made significant
contributions to the advancements in the quantum
field~\cite{Arkhipov-encircling2,wu-encircling,bell-encircling,Hassan-encircling2,Zhang-encircling,Nasari-encircling}.

However, it is regrettable that the existing
research~\cite{bell-encircling,topology-encircling,Arkhipov-encircling2,wu-encircling,Berry-encircling,Xu-encircling,Li-encircling,Arkhipov-encircling1,Ergoktas-encircling,Tang-encircling,
Hassan-encircling,Hassan-encircling2,Zhang-encircling,Nasari-encircling,Feilhauer-encircling}
predominantly focuses on either symmetric or asymmetric state
transfer, with limited exploration of simultaneous involvement of
both types or mode conversion between the two within the same
quantum system. Furthermore, the current schemes for state
transfer and their underlying protocols critically rely on the
presence of exceptional points (EPs), requiring specific designs
of non-Hermitian effective Hamiltonians with finely tuned
parameters. Moreover, these schemes also impose constraints
primarily associated with dissipation, typically considering only
a single type such as spontaneous radiation dissipation or cavity
decay~\cite{Arkhipov-encircling2,wu-encircling,open1,open2}, which
differs from real-world scenarios where quantum systems often
experience multiple forms of dissipation. Finally, while achieving
successful demonstrations of symmetric or asymmetric state
transfer in single-bit systems is noteworthy, efficiently
accomplishing entangled state transfers (i.e., Bell
states~\cite{Bellstate1,Bellstate2}) in systems with multiple
types of dissipation remains a challenging task.

In this work, we demonstrate  efficient and controlled symmetric
and asymmetric Bell-state transfers through modulation of system
parameters in a Jaynes-Cummings (J-C) model with both atomic
spontaneous emission and cavity decay. We effectively mitigate the
influence of nonadiabatic transitions induced by the imaginary
component of the eigenenergy spectrum through appropriate
parameter configurations. Subsequently, by selecting a trajectory
that dynamically encircles an EP, we successfully realize a
long-desired symmetric Bell-state transfer, wherein the exchange
of Bell states occurs dynamically irrespective of the encircling
direction. Moreover, we present a counterintuitive finding, which
suggests that the presence of an EP may not be indispensable for
the successful implementation of asymmetric state transfer in NH
systems. Furthermore, we achieve perfect asymmetric Bell-state
transfers even in the absence of an EP, while dynamically orbiting
around an approximate EP (AEP). Finally, we demonstrate that our
results hold for the J-C model under time-modulated and
time-independent dissipations, where the final Bell-state transfer
is determined by the trajectory orientation. Our work presents a
novel approach for effectively and reliably manipulating entangled
states with both symmetric and asymmetric characteristics through
the dissipation engineering in NH systems.

The rest of this paper are organized as follows. In
Sec.~\ref{section:II}, we provide a concise overview of the
dissipative Jaynes-Cummings model. Sections~\ref{section:III}
and~\ref{section:IV} are dedicated to detailed analyses of the
energy spectrum, eigenstate population distribution for symmetric
and asymmetric Bell-state transfer, respectively.  Finally, we
present a summary in Section~\ref{section:V}.

\section{Dissipative Jaynes-Cummings model}\label{section:II}

We consider a dissipative Jaynes-Cummings (J-C) model, in which a
two-level atom (with the ground state $|g\rangle$ and the excited
state $|e\rangle$) is coupled to an optical cavity with the
coupling strength $g$. The decay rate of the cavity mode and the
atomic spontaneous emission rate are respectively indicated by
$\kappa$ and $\gamma$. The effective Hamiltonian describing this
system can be expressed as follows~\cite{JC1,JC2,JC3,JC4} (taking
$\hbar$=1)
\begin{eqnarray}\label{eq1-1}
H_{\textrm{eff}}&=&\omega_{a}a^{\dag}a+\omega_{e}\sigma_{+}\sigma_{-}+g(a^{\dag}\sigma_{-}+a\sigma_{+})\cr
&&-\frac{i\kappa}{2}a^{\dag}a-\frac{i\gamma}{2}\sigma_{+}\sigma_{-},
\end{eqnarray}
where $\omega_{a}$ is the cavity resonance frequency, while
$\omega_{e}$ is the transition frequency between the two levels
$|g\rangle$ and $|e\rangle$ of the atom. The creation
(annihilation) operator of the cavity mode is denoted as
$a^{\dag}$ $(a)$, and the raising (lowering) operator of the atom
is represented by $\sigma_{+}$ $(\sigma_{-})$.

In the absence of an additional driving field, we make a
reasonable assumption that the total excitation of the system is
unity~\cite{JC2,Singleexcitation1,experiment-realize4,experiment-realize5,experiment-realize6}.
We denote the vacuum state (the singe photon state) of the cavity
as $|0\rangle$ ($|1\rangle$). In the basis states
$|e,0\rangle$=$|1,0\rangle^{T}$ and
$|g,1\rangle$=$|0,1\rangle^{T}$, where \textit{T} stands for
transpose, the effective Hamiltonian can be expressed as
\begin{eqnarray}\label{eq-Heff}
H_{\textrm{eff}}^{1}=\left(\begin{array}{ccccccc}
\omega_{a}+\delta-\frac{i\gamma}{2} &  g\\
g & \omega_{a}-\frac{i\kappa}{2} \\
\end{array}\right),
\end{eqnarray}
where $\delta$=$\omega_{e}-\omega_{a}$ is the difference between
the atomic transition frequency and the cavity frequency. The
eigenvalues of the NH Hamiltonian ($H_{\textrm{eff}}^{1}$) are
\begin{eqnarray}\label{eq1-3}
E_{\pm}&=&\frac{1}{4}[2(2\omega_{a}+\delta)-i(\gamma+\kappa)\pm{2\Delta_E}],\cr
\Delta_E&=&\frac{i}{2}\sqrt{(\gamma-\kappa+2i\delta)^2-16g^{2}},
\end{eqnarray}
and the corresponding right eigenvectors  are
\begin{eqnarray}\label{eq1-4}
|\phi_{+}\rangle&=&\frac{1}{S_{+}}[A_{+},4g]^{T},\cr
|\phi_{-}\rangle&=&\frac{1}{S_{-}}[A_{-},4g]^{T},
\end{eqnarray}
where $A_{\pm}=2\delta-i(\gamma-\kappa)\pm{2\Delta_E}$ and
$S_{\pm}=\sqrt{|A_{\pm}|^2+|4g|^2}$,  together with the left
eigenvectors
\begin{eqnarray}\label{eq1-5}
\langle\widehat{\phi_{+}}|&=&\frac{1}{S_{+}}[A_{+},4g],\cr
\langle\widehat{\phi_{-}}|&=&\frac{1}{S_{-}}[A_{-},4g].
\end{eqnarray}
The biorthogonal partners $\{\langle\widehat{{\phi_{n}}}|\}$,
$\{|{\phi}_{m}\rangle \}$ $(n,m$=$+,-)$ are normalized to satisfy
the biorthogonality and relations
$\langle\widehat{{\phi_{n}}}|\phi_{m}\rangle$=$\delta_{nm}$,
$\sum_{n}|\widehat{{\phi_{n}}}\rangle\langle\phi_{n}|$=$\sum_{n}|{\phi_{n}}\rangle\langle\widehat{{\phi_{n}}}|$=1~\cite{NH3}.

The difference $\Delta_E(\delta,g,\gamma,\kappa)$ between the
eigenvalues ($E_{\pm}$) plays a pivotal role in determining the
eigenenergy spectrum and the system dynamics. The analysis of
Eq.~(\ref{eq1-3}) reveals that an EP occurs at
$\delta=g=\gamma-\kappa=0$ when
$\Delta_E(\delta,g,\gamma,\kappa)=0$. However, analyzing the
relationship between all four parameters
$(\delta,g,\gamma,\kappa)$ and $\Delta_E$ simultaneously presents
a significant challenge. Therefore, it is recommended to fix
certain parameters and thoroughly investigate the correlation
between the remaining parameters and $\Delta_E$. This step is
crucial for identifying EPs within the system.

Once the EPs are determined, the primary objective is to design a
time evolution trajectory that encompasses the EPs in order to
achieve asymmetric or symmetric Bell-state transfer. In this
context, we provide several remarks regarding the design of the
time evolution trajectory. (i) It is crucial to ensure that the
eigenvectors $|\phi_{+}\rangle$ and $|\phi_{-}\rangle$ correspond
to Bell states $(|1,0\rangle\pm|0,1\rangle)/\sqrt{2}$ at both the
beginning and end of the trajectory.  To ensure the formation of
Bell states, it is necessary to maintain
$\delta_{t_i,t_f}$=$(\gamma-\kappa)_{t_i,t_f}$=$0$ [where $t_i$
($t_f$) represents the initial (final) time], thereby
$|\phi_{+}(t_i,t_f)\rangle$ and $|\phi_{-}(t_i,t_f)\rangle$
becoming Bell states can be satisfied. (ii) Both the real and
imaginary components of the time evolution trajectory should be
carefully designed to encircle or closely approach the EP.  The
subsequent sections will be dedicated to the exploration of
appropriate time evolution trajectories for achieving symmetric
and asymmetric Bell-state transfers.

\section{symmetric Bell-state transfer}\label{section:III}

Recent
studies~\cite{topology-encircling,Arkhipov-encircling2,wu-encircling}
have demonstrated that achieving efficient symmetric state
transfers hinges on satisfying the generalized adiabatic condition
by effectively suppressing nonadiabatic couplings between
instantaneous eigenstates. In this section, we propose a
controlled scheme to realize a long-sought symmetric Bell-state
transfer, where the Bell-state transfer occurs dynamically
regardless of the encircling direction. This is accomplished by
slowing down the system evolution and minimizing the imaginary
component
$\textrm{Im}[\Delta_E(\delta,g,\gamma,\kappa)]$~\cite{wu-encircling}.
Specifically, the imaginary component
$\textrm{Im}[\Delta_E(\delta,g,\gamma,\kappa)]$ can be effectively
eliminated by setting appropriate parameter configurations, such
as $\delta=0$ and $16g^2 > (\gamma-\kappa)^2$. For the sake of
discussion, we assume $\kappa=\alpha\gamma$ and normalize all
parameters with respect to the cavity frequency $\omega_a$ in the
subsequent analysis.

\begin{figure}[htb]
\scalebox{0.4}{\includegraphics{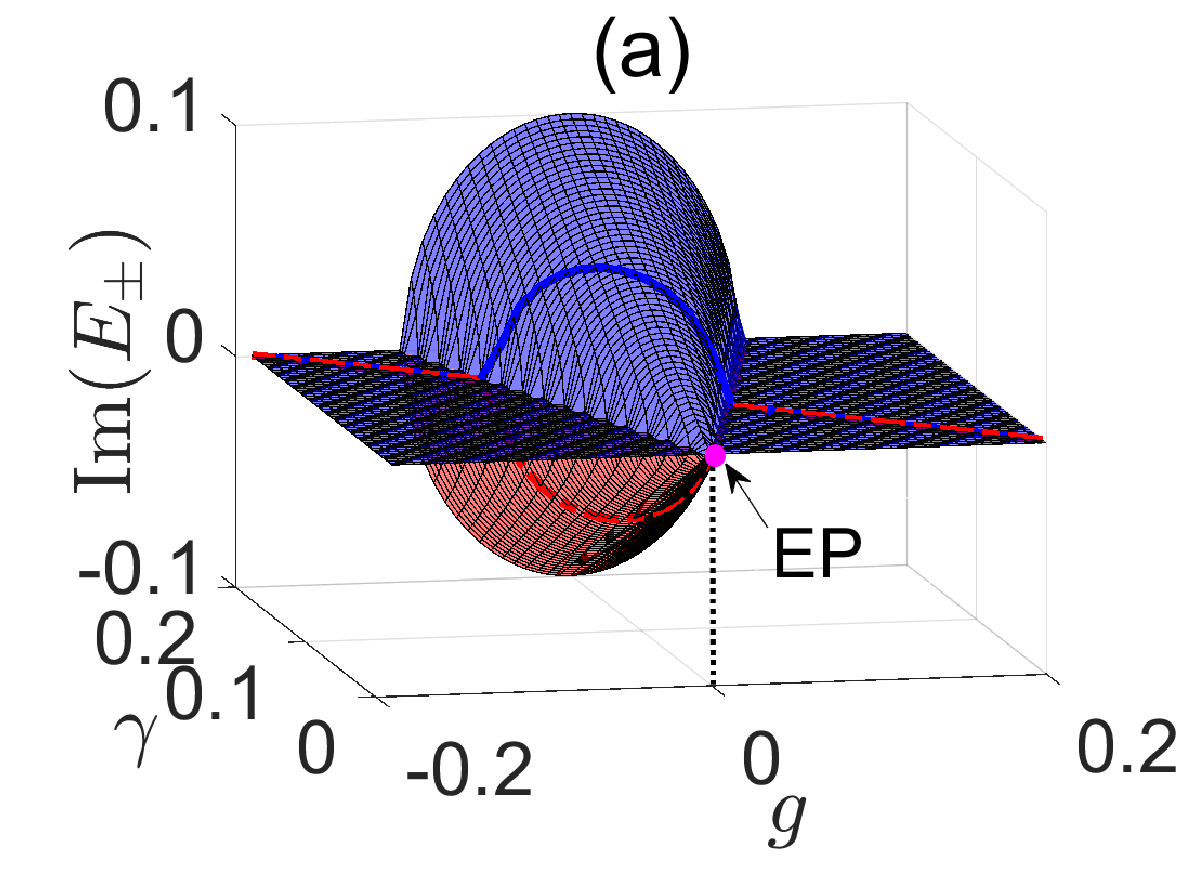}}
\scalebox{0.4}{\includegraphics{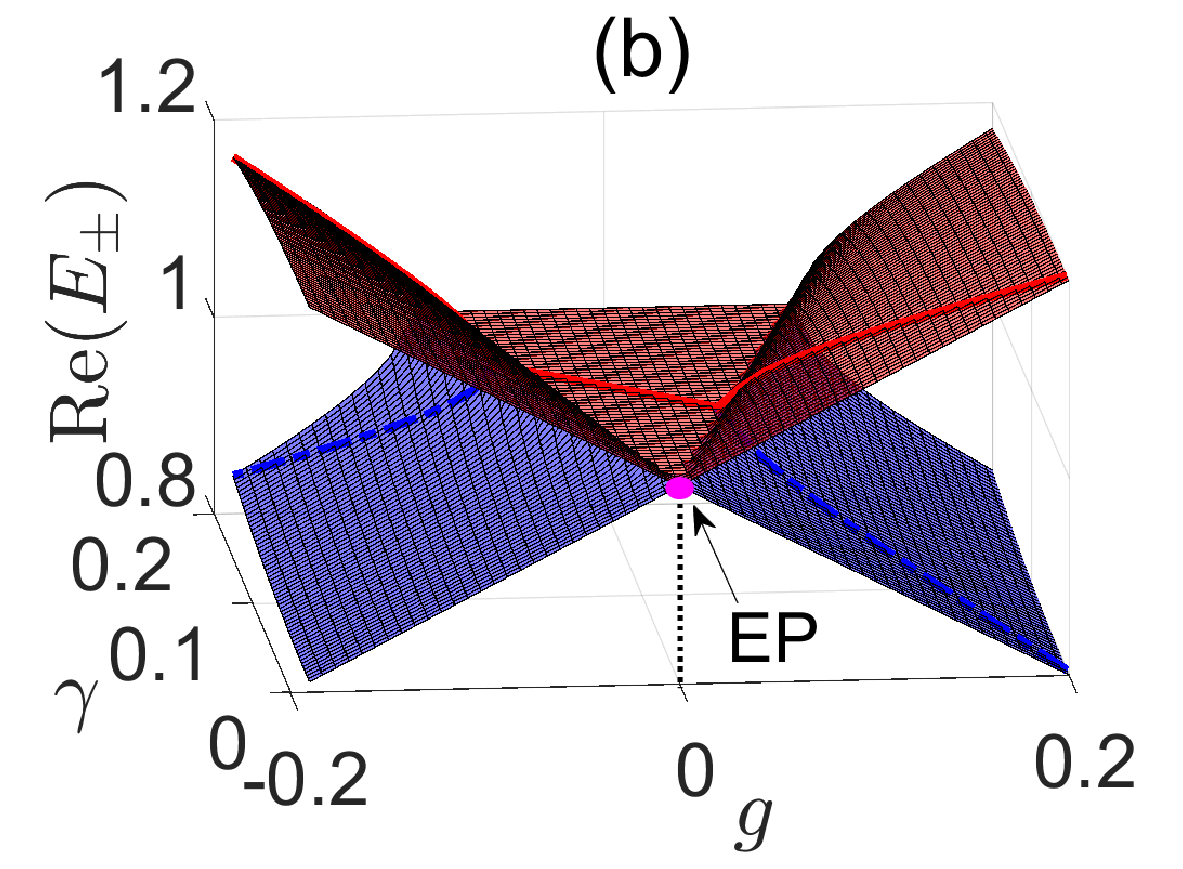}}
\caption{\label{fig-E-symmetric} The eigenenergy spectrum
$E_{\pm}(g,\gamma)$ and the system's time-evolution trajectory
$E_{\pm}[g(t),\gamma(t)]$ for symmetric Bell-state transfer. (a)
the  imaginary  part and (b) the real part of $E_{\pm}(g,\gamma)$
and $E_{\pm}[g(t),\gamma(t)]$. The blue (red) Riemann surface
corresponds to $E_{+}(g,\gamma)$ [$E_{-}(g,\gamma)$], and the
solid blue  (dashed red) line represents $E_{+}[g(t),\gamma(t)]$
($E_{-}[g(t),\gamma(t)])$, respectively. The pink point is an EP.
The other parameters are chosen as $g_{0}=0.01,$
$G_{0}=\Gamma_{0}=0.2,$ $\alpha=-1$, and $\omega=\pi$.}
\end{figure}

Based on the aforementioned settings, we can investigate the
intricate correlation between the energy difference and the
remaining two parameters $(g,\gamma)$ in a three-dimensional
context. In order to gain an intuitive understanding of the energy
variation within the system, in Fig.~\ref{fig-E-symmetric}, we
illustrate the eigenenergy spectrum $E_{\pm}$ in the parameter
space $(g,\gamma)$.  From Fig.~\ref{fig-E-symmetric}, it is
evident that the imaginary parts of $E_+$ and $E_-$ intersect with
the plane defined by $E=0$, their real counterparts intersect with
another plane characterized by $E=1$. The intersection point of
these two planes within the parameter space $(g,\gamma)$
corresponds to a pink point at coordinates (0,0), which represents
an EP. Consequently, by steering our system's evolution trajectory
around or in close proximity to this EP, we can achieve a
symmetric transfer of Bell states as desired.

A symmetric Bell-state transfer can be realized through the
careful selection of the parameter trajectory
\begin{eqnarray}\label{eq1-6}
g(t)=g_{0}+G_{0}\cos(\omega{t}),\gamma(t)=\Gamma_{0}\sin^2(\omega{t}),
\end{eqnarray}
where  $g_{0}$, $G_{0}$, $\Gamma_{0}$, and $\omega$ are real
constants.  According to Eq.~(\ref{eq1-6}), the NH Hamiltonian
$H_{eff}^{1}(t)$ exhibits periodic variations with respect to time
$t$, having a period of $T=\pi/\omega$.
Figure~\ref{fig-E-symmetric} also illustrates the time-evolution
trajectories, where the solid blue line represents
$E_{+}[g(t),\gamma(t)]$ and the dashed red line corresponds to
$E_{-}[g(t),\gamma(t)]$. It can be observed that both trajectories
orbit in the vicinity of the pink EP, in accordance with our
assumptions.

\begin{figure}
\scalebox{0.42}{\includegraphics{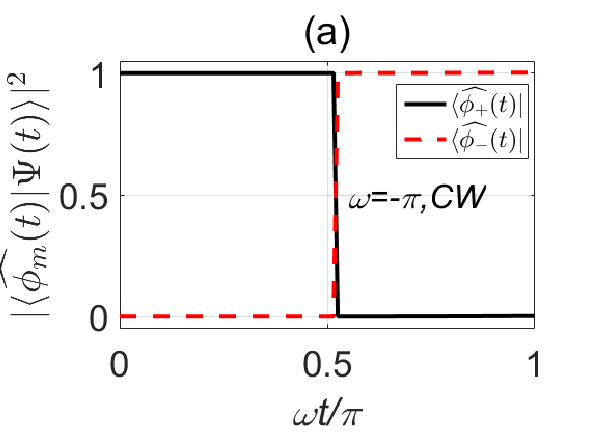}\includegraphics{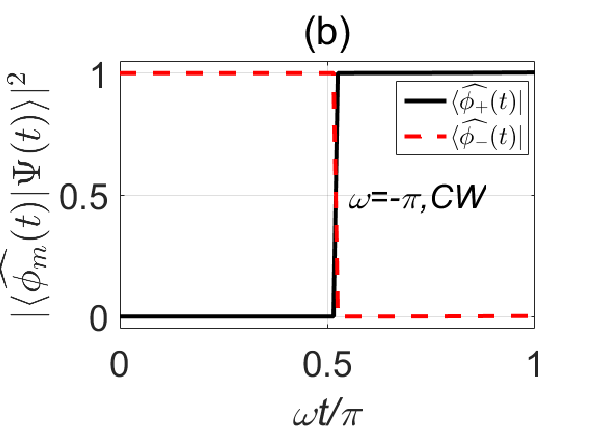}}
\scalebox{0.42}{\includegraphics{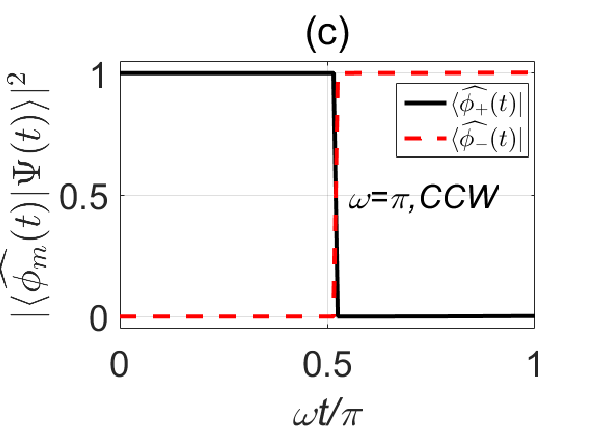}\includegraphics{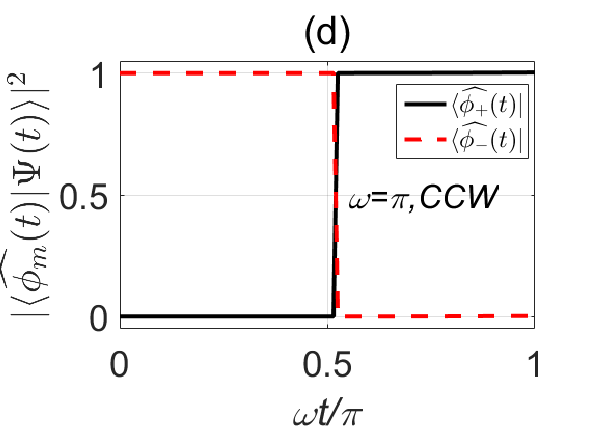}}
\caption{\label{fig-F-symmetric}   The time evolution of the
fidelity $F_{m}=|\langle\widehat{{\phi_{m}}}(t)|\Psi(t)\rangle|^2$
for the time-depend right eigenstate $|\phi_{n}(t)\rangle$
$(m=+,-)$. The initialized state in (a) and (c) [(b) and (d)] is
chosen as $|\Psi(0)\rangle$=$|\phi_{+}(0)\rangle$
($|\phi_{-}(0)\rangle$). The direction of trajectory  in (a)-(b)
is clockwise (CW)  and in (c)-(d) is counterclockwise (CCW). The
other parameters are chosen as $g_{0}$=0.01,
$G_{0}$=$\Gamma_{0}$=0.2, and $\alpha$=-1. The state dynamics
exhibits a purely adiabatic character, enabling one to implement a
symmetric Bell-state switch. }
\end{figure}

To evaluate the validity of the symmetric Bell-state transfer, we
investigate the dynamics of the system. The fidelity of the right
eigenstate $|\phi_{m}(t)\rangle$ ($m$=$+,-$) is determined by the
relation
$F_{m}=|\langle\widehat{{\phi_{m}}}(t)|\Psi(t)\rangle|^2$, where
$|\Psi(t)\rangle$ represents the evolving state of the system at
time $t$. Assuming that the system is initially prepared in one of
its right eigenstates, i.e.,
$|\Psi(0)\rangle$=$|\phi_{\pm}(0)\rangle$=$(|e,0\rangle\pm|g,1\rangle)/\sqrt{2}$,
we numerically integrate Schr\"{o}dinger's equation
\begin{eqnarray}\label{eq1-13}
i\partial_{t}|\Psi(t)\rangle=H_{\textrm{eff}}^{1}|\Psi(t)\rangle,
\end{eqnarray}
to obtain the evolving state $|\Psi(t)\rangle$. The instantaneous
left eigenvector $\langle\widehat{{\phi_{m}}}(t)|$ can be
calculated by substituting parameters into Eqs.~(\ref{eq1-5}) and
(\ref{eq1-6}). The results of these calculations for different
encircling directions and initialized states are illustrated in
Fig.~\ref{fig-F-symmetric}. Figures~\ref{fig-F-symmetric}(a)-(d)
demonstrate complete Bell-state transfers (achieving maximal
transfer fidelity 1 by employing sufficiently slow trajectories),
where $|\phi_{-}(t)\rangle$$\leftrightarrow$$|\phi_{+}(t)\rangle$
are exchanged after one period $T=\pi/\omega$, irrespective of
encircling direction. The state evolution in this context
demonstrates exclusive adiabatic characteristics, facilitating the
implementation of asymmetric Bell-state switching.

\section{asymmetric Bell-state transfer}\label{section:IV}

\subsection{Asymmetric Bell-state transfer under time-modulated dissipations}
\begin{figure}
\scalebox{0.32}{\includegraphics{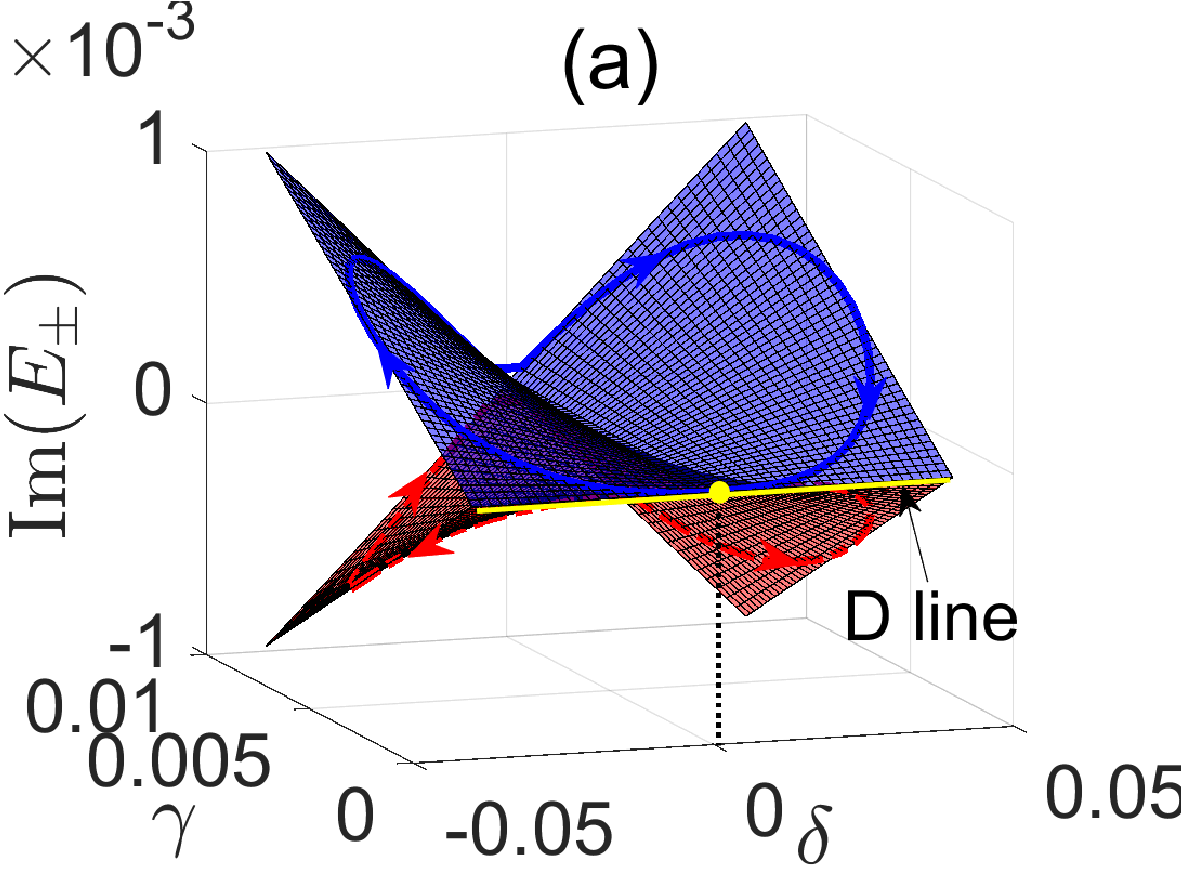}}
\scalebox{0.32}{\includegraphics{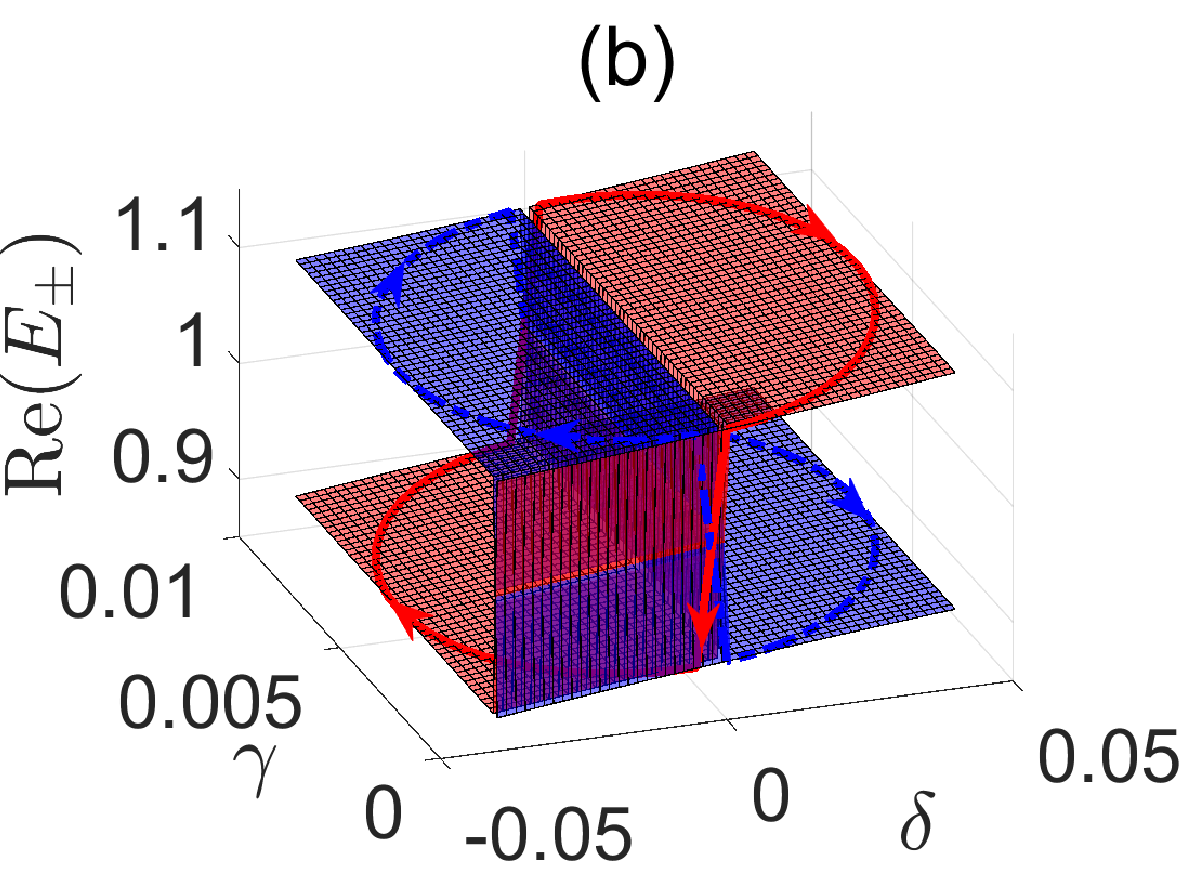}}
\scalebox{0.32}{\includegraphics{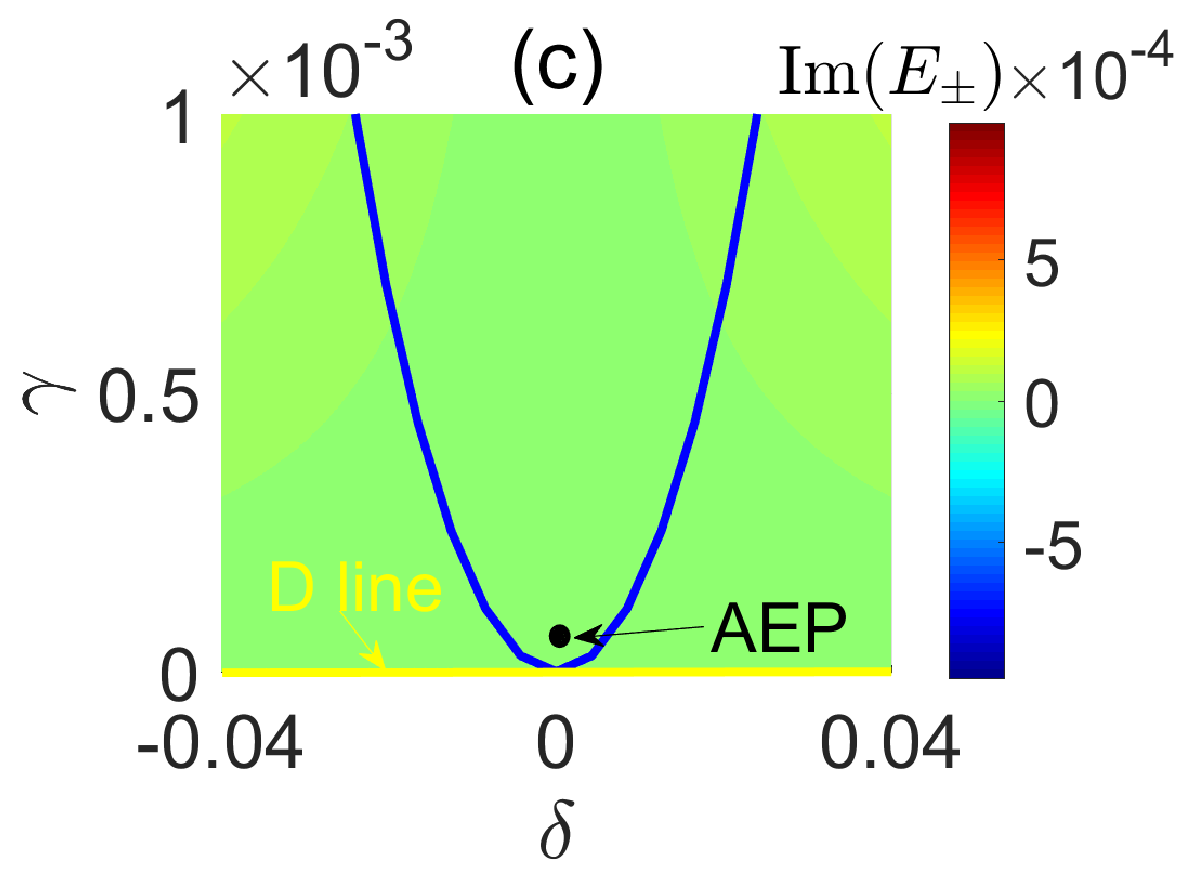}}
\scalebox{0.32}{\includegraphics{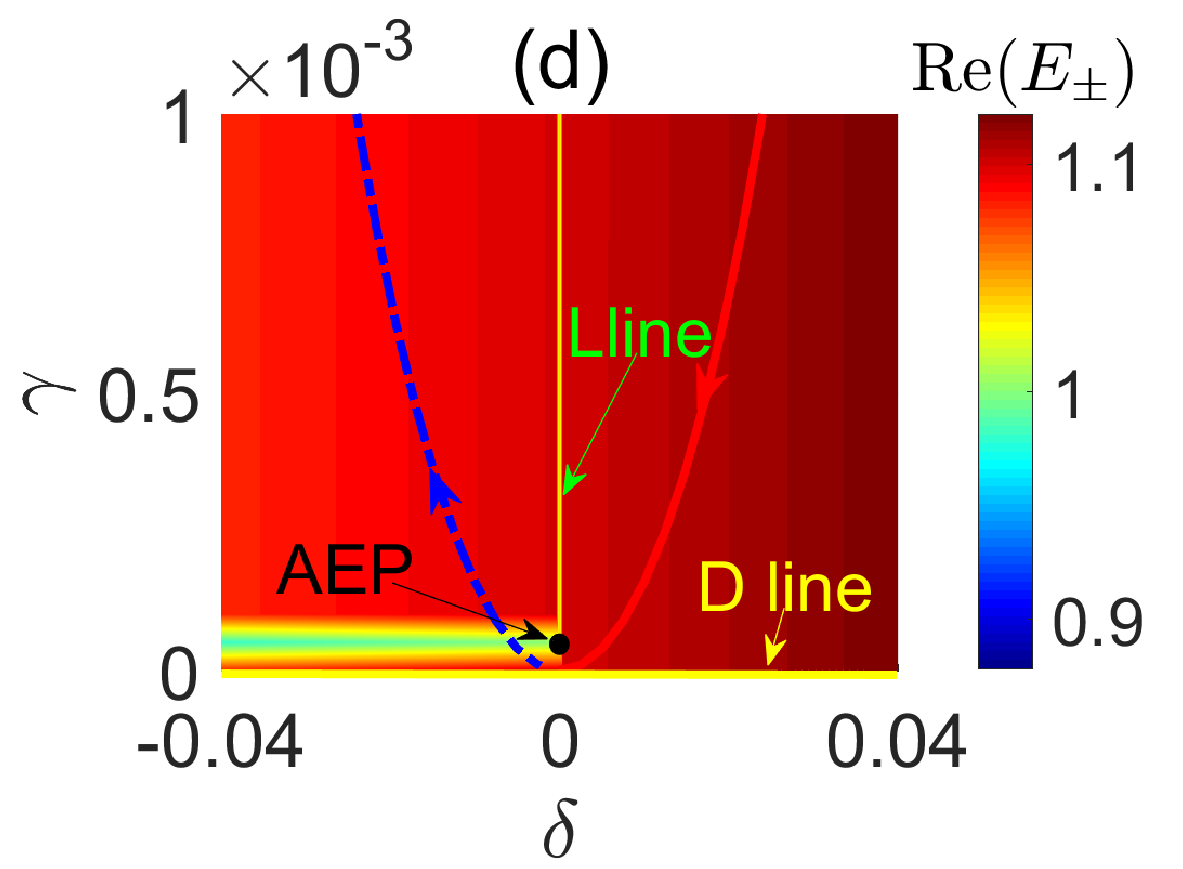}}
\caption{\label{fig-E-asymmetric-im} The eigenenergy spectrum
$E_{\pm}(\gamma,\delta)$ and the system's time-evolution
trajectory $E_{\pm}[\gamma(t),\delta(t)]$ for a asymmetric
Bell-state transfer. The imaginary and real parts of the
eigenenergy spectrum are depicted in panels (a) and (b) as
three-dimensional surfaces, while in panels (c) and (d), they are
presented as two-dimensional plots by fixing $\gamma \in (0,
0.001)$. The D line and L line represent the intersection lines of
the surfaces corresponding to the imaginary and real parts of the
eigenenergy spectrum $E_{\pm}(\gamma,\delta)$, respectively. The
system's time-evolution trajectory is dynamically orbiting around
an AEP, the black point $(0, 1 \times 10^{-4})$, in the parameter
space $(\gamma,\delta)$. The other parameters are chosen as
$g_{0}=0.1$, $\Delta_{0}=0.04$, $\Gamma_{0}=0.1$, and $\alpha=-1$,
and $\omega=\pi$. }
\end{figure}

Generally, the successful transfer of asymmetric or chiral states
critically depends on the topological properties of the
intersecting Riemann energy manifolds at
EPs~\cite{bell-encircling,Berry-encircling,Xu-encircling,Li-encircling,Arkhipov-encircling1,Ergoktas-encircling,Tang-encircling,Hassan-encircling,Feilhauer-encircling}.
Here, we propose a controlled scheme to realize the asymmetric
Bell-state transfer. Specifically, we begin by fixing the value of
$g$ and systematically analyzing the relationship between the
energy difference and the parameters ($\delta$, $\gamma$).
Additionally, we present a graphical depiction of the eigenenergy
spectrum $E_{\pm}$ in the parameter space ($\delta$, $\gamma$), as
illustrated in Fig.~\ref{fig-E-asymmetric-im}.
Figures~\ref{fig-E-asymmetric-im}(a) and (b) respectively display
the three-dimensional variations of the imaginary and real parts
of the eigenenergy spectrum $E_{\pm}(\gamma,\delta)$, as well as
the time-evolution trajectory of the system
$E_{\pm}[\gamma(t),\delta(t)]$. Furthermore,
Figs.~\ref{fig-E-asymmetric-im}(a) and (c) reveal that the
imaginary components of $E_+(\gamma,\delta)$ and
$E_-(\gamma,\delta)$ intersect along the diabolic line (D Line),
where $\textrm{Im}(E_{+}) = \textrm{Im}(E_{-}) = 0$. Meanwhile,
Figs.~\ref{fig-E-asymmetric-im}(b) and (d) demonstrate that the
real components of $E_+(\gamma,\delta)$ and $E_-(\gamma,\delta)$
converge along an L-shaped curve (L Line), which corresponds to
$\textrm{Re}(E_{+})=\textrm{Re}(E_{-})=1$. However, by comparing
Figs.~\ref{fig-E-asymmetric-im}(c) and (d), it becomes apparent
that the imaginary and real parts of the spectrum
$E_{\pm}(\gamma,\delta)$ cannot coincide simultaneously
[$\textrm{Im}(E_{+})=\textrm{Im}(E_{-})$ and
$\textrm{Re}(E_{+})=\textrm{Re}(E_{-})$], as evidenced by the
misalignment between the D line and the L line. This observation
conclusively verifies the absence of an exceptional point
($E_{+}=E_{-}$) under the current parameter settings.

\begin{figure}
\scalebox{0.42}{\includegraphics{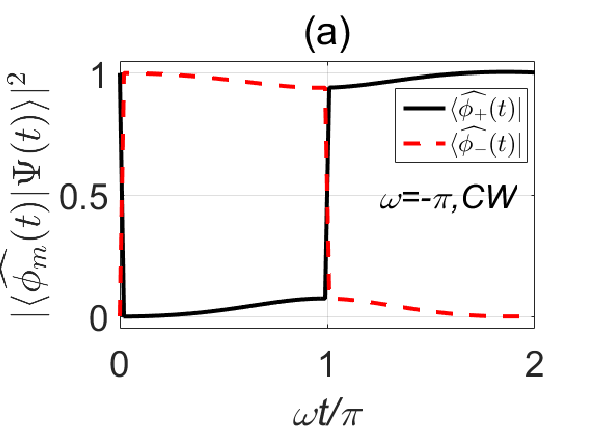}\includegraphics{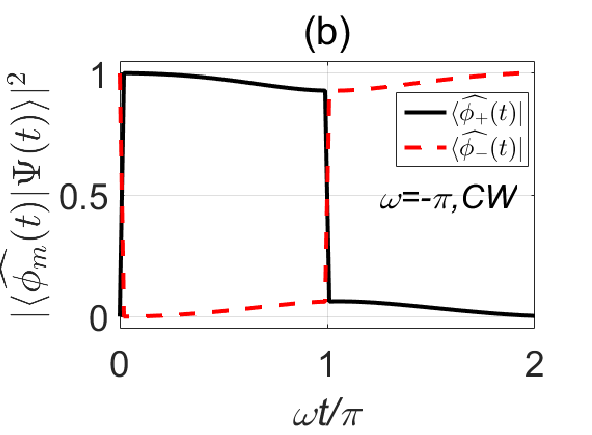}}
\scalebox{0.42}{\includegraphics{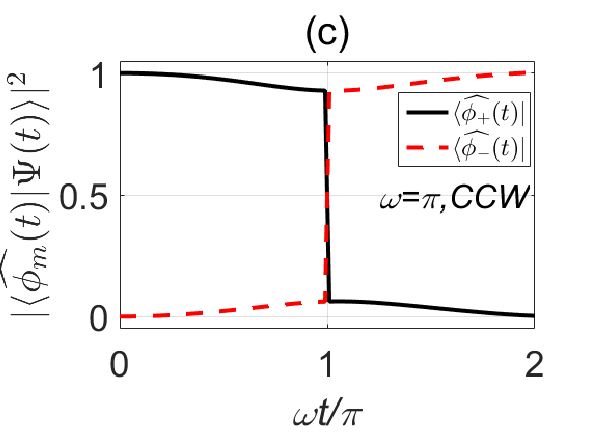}\includegraphics{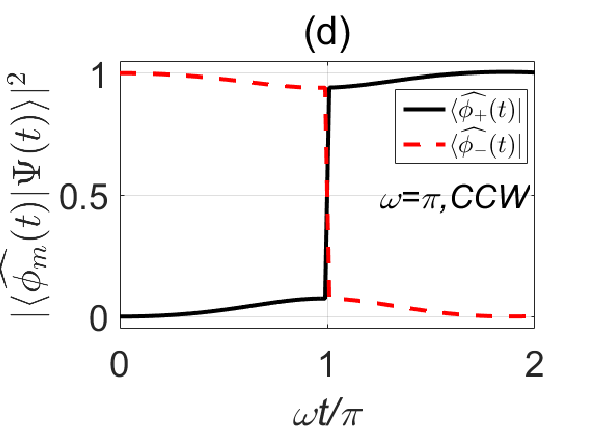}}
\caption{\label{fig-F-asymmetric-im}   The time evolution of the
fidelity $F_{m}=|\langle\widehat{{\phi_{m}}}(t)|\Psi(t)\rangle|^2$
for the time-depend right eigenstate $|\phi_{m}(t)\rangle$
$(m=+,-)$. The initialized state in (a) and (c) [(b) and (d)] is
chosen as $|\Psi(0)\rangle$=$|\phi_{+}(0)\rangle$
($|\phi_{-}(0)\rangle$). The selected trajectories in (a)-(b)
exhibit a clockwise (CW) encircling, while those in (c)-(d)
demonstrate a counterclockwise (CCW) encircling. Other parameters
are set as follows: $g_{0}=0.1$, $\Delta_{0}=0.04$,
$\Gamma_{0}=0.1$, and $\alpha=-1$.  The state dynamics exhibits a
chiral in nature, thus, enabling one to implement an asymmetric
Bell-state switch.}
\end{figure}

According to the conventional notion, the presence of EPs is
deemed a prerequisite for the viability of the scheme, suggesting
that chiral transformations are unattainable in their absence.
However, it is intriguing to unveil that even when the system
evolves around an AEP, a significant chiral transition still
endures. This feasibility stems from the Riemann topology of the
system spectrum induced by AEP and a specific selection of
encircling trajectory within the parameter space. For example, an
asymmetric Bell-state transfer can be attained when the system's
time-evolution trajectory is dynamically orbiting around an AEP
(see the black point in Fig.~\ref{fig-E-asymmetric-im}), in the
parameter space $(\gamma,\delta)$. Without loss of generality, the
system's time-evolution trajectory can be set as
\begin{eqnarray}\label{eq1-7}
\delta(t)=\Delta_{0}\sin(\omega{t}),\gamma(t)=\Gamma_{0}\sin^2(\frac{\omega{t}}{2}),
\end{eqnarray}
where $\Delta_{0},\Gamma_{0},$ and $\omega \in \mathbf{R}$ are
constants, the period of the trajectory is $T=2\pi/\omega$. The
time-evolution trajectories  $E_{\pm}[\gamma(t),\delta(t)]$ are
depicted in Fig~\ref{fig-E-asymmetric-im}.

In Fig.~\ref{fig-F-asymmetric-im}, we present the temporal
evolution of the fidelity for the right eigenstate
$|\phi_{m}(t)\rangle$ $(m=+,-)$ under different initial states and
encircling directions. The orbiting trajectory proceeds in a
counterclockwise (CCW) or clockwise (CW) direction for positive
angular frequencies $\omega>0$ or negative angular frequencies
$\omega<0$, respectively. As illustrated in
Figs.~\ref{fig-F-asymmetric-im}(a) and (b), the right eigenstates
(Bell states) return to their original configurations after
completing a dynamical cycle with a CW encircling direction.
However, as shown in Figs.~\ref{fig-F-asymmetric-im}(c) and (d),
the Bell states undergo an exchange after completing a dynamical
cycle with a CCW encircling direction. The state dynamics exhibits
a distinct chiral character, facilitating the implementation of a
symmetric Bell-state switch.

\subsection{Asymmetric Bell-state transfer under time-independent dissipations}

\begin{figure}[htb]
\scalebox{0.4}{\includegraphics{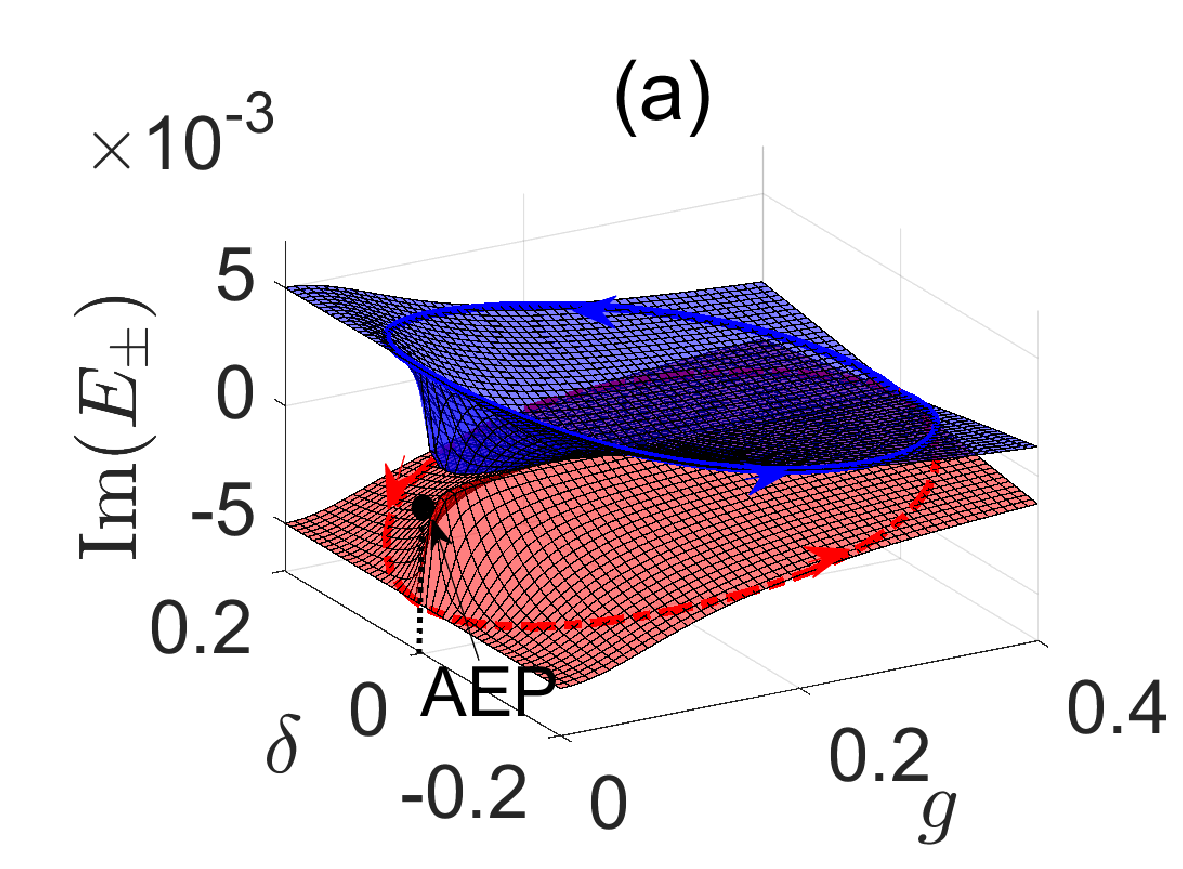}}
\scalebox{0.4}{\includegraphics{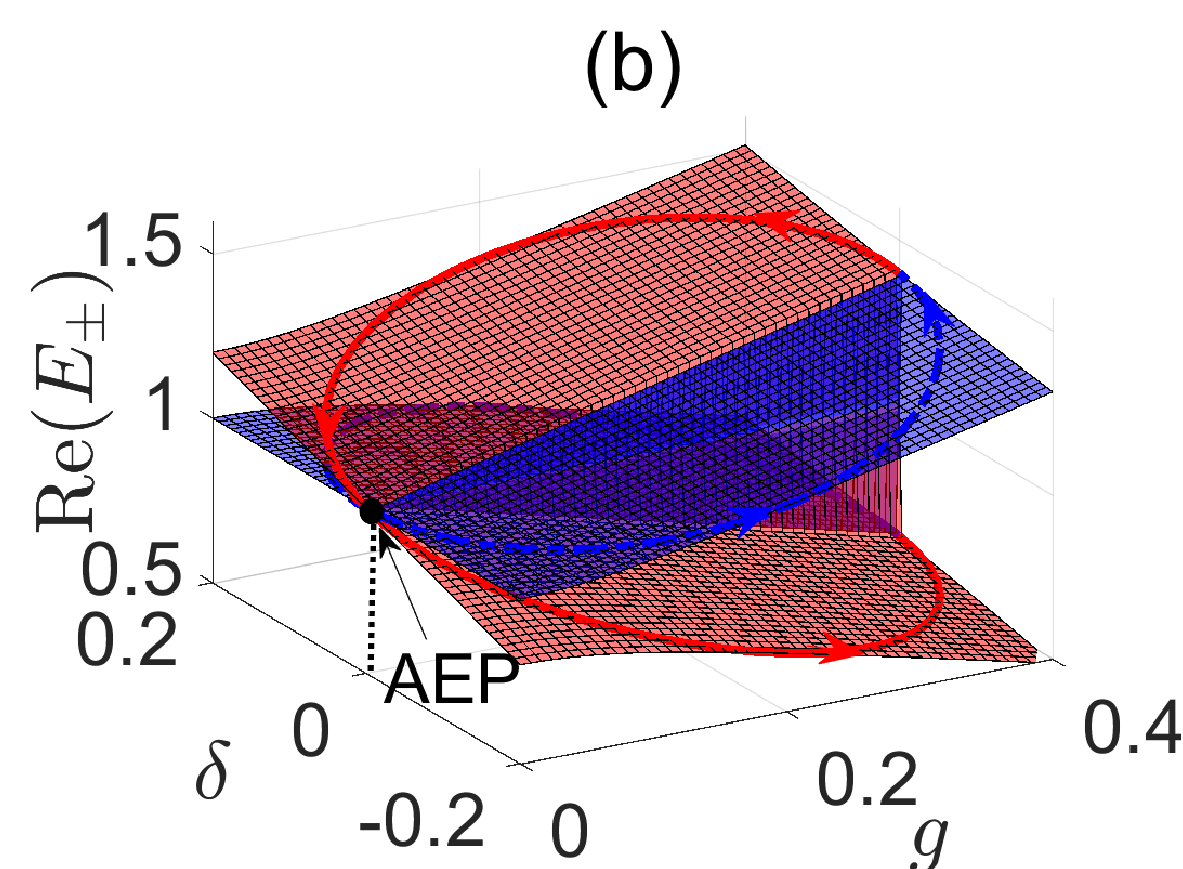}}
\caption{\label{fig-E-asymmetric-c} The eigenenergy spectrum
$E_{\pm}(g,\delta)$ and the system's time-evolution trajectory
$E_{\pm}[g(t),\delta(t)]$ for a dissipative system with
time-independent dissipation. (a) the real part and (b) the
imaginary part of $E_{\pm}(g,\gamma)$ and
$E_{\pm}[g(t),\gamma(t)]$. The red (blue) Riemann surface
corresponds to $E_{+}(g,\gamma)$ [$E_{-}(g,\gamma)$], and the
solid red (dashed blue) line represents $E_{+}(g,\gamma)$
($E_{-}(g,\gamma))$, respectively. The pink point is an AEP. Other
parameters are set as follows: $g_{0}=G_{0}=\Delta_{0}=0.2$,
$\gamma_{0}=0.1$, $\alpha=-1$, and $\omega=\pi$.}
\end{figure}

We have successfully demonstrated the remarkable achievement of
chiral Bell-state transfer through modulating time-dependent
dissipative parameters. However, it is important to acknowledge
that managing noise parameters can present challenges in certain
systems or experiments. From an experimental perspective, we
propose an alternative approach for achieving chiral Bell-state
transfer by utilizing time-independent dissipative parameters.
Furthermore, we provide the eigenenergy spectrum $E_{\pm}$  in the
parameter space $(g,\delta)$ as shown in
Fig~\ref{fig-E-asymmetric-c}. Figure~\ref{fig-E-asymmetric-c}(a)
demonstrates the separation between the imaginary parts of
$E_\pm(g,\delta)$ at point (0,0), while
Fig~\ref{fig-E-asymmetric-c}(b) showcases the convergence of the
real parts of $E_\pm(g,\delta)$ at point (0,0). Hence, under
current parameter configuration, the system does not manifest any
EP; instead, it exhibits a distinctive black AEP.

\begin{figure}
\scalebox{0.42}{\includegraphics{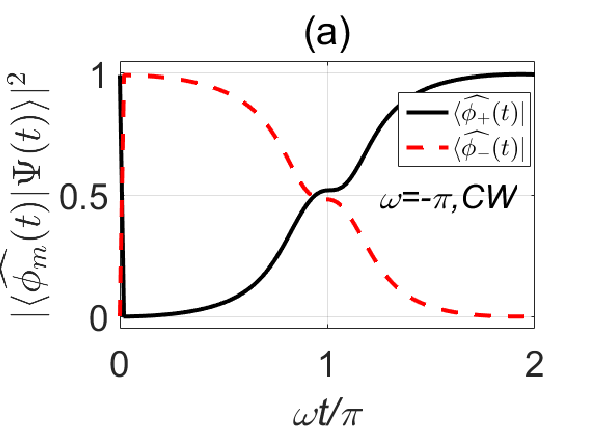}\includegraphics{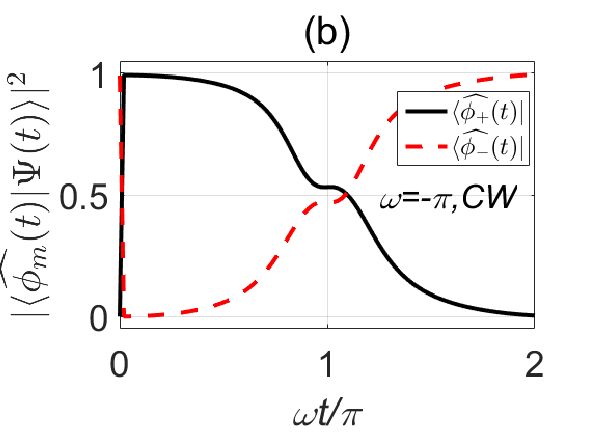}}
\scalebox{0.42}{\includegraphics{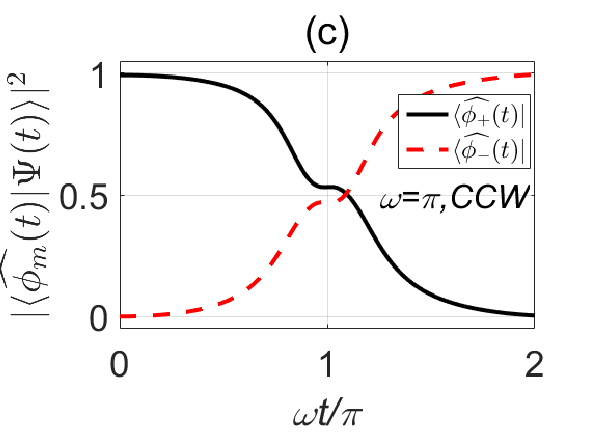}\includegraphics{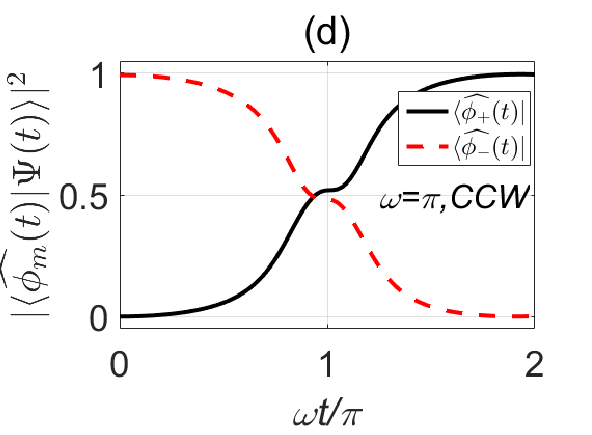}}
\caption{\label{fig-F-asymmetric-c}  The time evolution of the
fidelity $F_{m}=|\langle\widehat{{\phi_{m}}}(t)|\Psi(t)\rangle|^2$
is examined for the time-dependent right eigenstate
$|\phi_{m}(t)\rangle$ $(m=+,-)$. The initial state in (a) and (c)
[(b) and (d)] is selected as $|\Psi(0)\rangle =
|\phi_{+}(0)\rangle$ ($|\phi_{-}(0)\rangle$). The chosen
trajectory in (a)-(b) exhibits a CW encircling, while that in
(c)-(d) displays a CCW encircling. The parameters are set as
follows: $g_{0}=G_{0}=\Delta_{0}=0.2$, $\gamma_{0}=0.1$, and
$\alpha=-1$.}
\end{figure}

The achievement of asymmetric Bell-state transfer can be realized
by carefully selecting the parameter trajectory as follows:
\begin{eqnarray}\label{eq1-8}
\gamma(t)=\gamma_{0},g(t)=g_{0}+G_{0}\cos(\omega{t}),\delta(t)=\Delta_{0}\sin(\omega{t}),
\end{eqnarray}
where $g_{0}, G_{0}, \gamma_{0}, \Delta_{0},$ and $\omega, \in
\mathbf{R}$ are constants, and the period of the trajectory is
$T=2\pi/\omega$. The time-evolution trajectories of $E_{\pm}(t)$
are also depicted in Fig~\ref{fig-F-asymmetric-c}. It can be
observed that the trajectories of $E_{\pm}[g(t),\gamma(t)]$
encircle around the AEP.  We also present the temporal evolution
of the fidelity for the right eigenstate $|\phi_{m}(t)\rangle$
$(m=+,-)$ under different initial states and encircling directions
in Fig.~\ref{fig-F-asymmetric-c}. The state dynamics also exhibits
a distinct chiral character, facilitating the implementation of a
symmetric Bell-state switch.

\section{CONCLUSION}\label{section:V}

We have theoretically demonstrated both symmetric and asymmetric
Bell-state transfers in a J-C model that incorporates atomic
spontaneous emission and cavity decay.  In the case of symmetric
Bell-state transfer, we have effectively mitigated the impact of
nonadiabatic transitions induced by the imaginary component of the
eigenenergy spectrum through appropriate parameter settings.
Thereby, we have achieved  an efficient transfer by selecting a
trajectory that dynamically encircles an EP. The proposed
procedure enables a long-desired symmetric switch, wherein the
Bell states are dynamically exchanged irrespective of the
encircling direction. Furthermore, we have demonstrated chiral
dynamics in the dissipative J-C model without a strict dependence
on exceptional points (EP). Specifically, we have realized
asymmetric Bell-state transfers even in the absence of EP while
dynamically orbiting around an approximate exceptional point
(AEP), where the final state is determined by trajectory
orientation. Lastly, we have proposed an alternative approach to
achieve chiral Bell-state transfers by utilizing time-independent
dissipative parameters.

The presented results have potential applications beyond the
Bell-state transfer through controlled dissipation, such as the
transfer of multi-mode (hybrid) entangled states (e.g., GHZ
state). Furthermore, our findings indicate that chiral dynamics
can be achieved in NH systems without strict EP conditions,
allowing for asymmetric state transfer while dynamically orbiting
an AEP. This result has immediate implications for reducing
parameter control complexity and experimental difficulties in NH
systems with multiple types of dissipation. Hence, our work
demonstrates a novel approach to manipulate entangled states with
both symmetric and asymmetric characteristics using dissipation
engineering, which is highly effective and reliable for
quantum-state engineering in non-Hermitian systems.

\section*{ACKNOWLEDGEMENT}
This work was supported by National Natural Science Foundation of
China (NSFC) (Grants Nos. 12264040, 12374333, and U21A20436),
National Key Research and Development Program of China (Grant No.
2024YFA1408900), Jiangxi Natural Science Foundation (Grant No.
20232BCJ23022), Innovation Program for Quantum Science and
Technology (Grant No.~2021ZD0301705) and the Jiangxi Province Key
Laboratory of Applied Optical Technology (Grant No.~2024SSY03051).


\begin{thebibliography}{999}

\bibitem{EPs1} L. Feng,  R. El-Ganainy and  L. Ge, ``Non-Hermitian photonics based on parity-time symmetry,'' Nat. Photon. \textbf{11}, 752 (2017).
\bibitem{EPs2}$\c{S}$. \"{O}zdemir,  S. Rotter, F. Nori, and L. Yang, ``Parity-time symmetry and exceptional points in photonics,'' Nat. Mater. \textbf{18}, 783 (2019).
\bibitem{EPs3}W. D. Heiss, ``Repulsion of resonance states and exceptional points,'' Phys. Rev. E \textbf{61}, 929 (2000).
\bibitem{EPs4}H. Cartarius, J. Main, and G. Wunner, ``Exceptional points in atomic spectra,'' Phys. Rev. Lett. \textbf{99}, 173003 (2007).
\bibitem{experiment-realize2}B. Peng, S. K. \"{O}zdemir, F. Lei, F. Monifi, M. Gianfreda, G. L. Long, S. Fan, F. Nori, C. M. Bender, and L. Yang, ``Parity-timesymmetric whispering-gallery microcavities,'' Nat. Phys. \textbf{10}, 394 (2014).

\bibitem{NH1}T. Kato, ``\textit{Perturbation Theory for Linear Operators},'' Classics in Mathematics (Springer, Berlin, 1995).
\bibitem{NH2}Y. Ashida, Z. Gong, and M. Ueda, ``Non-Hermitian physics,'' Adv. Phys. \textbf{69}, 249 (2020).
\bibitem{NH3} Q. C. Wu,  J. L. Zhao,  Y. L. Fang,  Y. Zhang,  D. X. Chen, C. P. Yang and F. Nori, ``Extension of Noether's theorem in PT-symmetry systems and its experimental demonstration in an optical setup,'' Sci. China-Phys. Mech. Astron. \textbf{66}(4), 240312 (2023).

\bibitem{sensitivity}H. Hodaei, A. U. Hassan, S. Wittek, H. Garcia-Gracia, R. El-Ganainy, D. N. Christodoulides, and M. Khajavikhan, ``Enhanced sensitivity at higher-order exceptional points,'' Nature \textbf{548}, 187 (2017).
\bibitem{invisibility}Z. Lin, H. Ramezani, T. Eichelkraut, T. Kottos, H. Cao, and D. N. Christodoulides, ``Unidirectional invisibility induced by PT-symmetric periodic structures,'' Phys. Rev. Lett. \textbf{106}, 213901 (2011).

\bibitem{lossinduced}A. Guo, G. J. Salamo, D. Duchesne, R. Morandotti, M. Volatier-Ravat, V. Aimez, G. A. Siviloglou, and D. N. Christodoulides, ``Observation of PT-symmetry breaking in complex optical potentials,'' Phys. Rev. Lett. \textbf{103}, 093902 (2009).

\bibitem{topology1} E. J. Bergholtz, J. C. Budich, and F. K. Kunst, ``Exceptional topology of non-Hermitian systems,'' Rev. Mod. Phys. \textbf{93}, 015005 (2021).
\bibitem{topology-encircling}C. Guria, Q. Zhong,  $\c{S}$. K. \"{O}zdemir, Y. S. S. Patil, R. El-Ganainy, and J. G. Emmet Harris, ``Resolving the topology of encircling multiple exceptional points,'' Nat. Commun. \textbf{15}, 1369 (2024).

\bibitem{Arkhipov-encircling2} I. I. Arkhipov, A. Miranowicz, F. Minganti, $\c{S}$. \"{O}zdemir, and F. Nori, ``Restoring adiabatic state transfer in time-modulated non-hermitian systems,'' Phys. Rev. Lett. \textbf{133}, 113802, (2024)
\bibitem{wu-encircling}Q. C. Wu,  J. L. Zhao, Y. H. Zhou, B. L. Ye, Y. L. Fang, Z. W., Zhou and C. P. Yang,  ``Shortcuts to adiabatic state transfer in time-modulated two-level non-Hermitian systems,'' Phys. Rev. A \textbf{111}, 022410 (2025).

\bibitem{bell-encircling}S. Khandelwal, W. J. Chen, K. W. Murch, and G. Haack, ``Chiral Bell-State Transfer via Dissipative Liouvillian Dynamics,'' Phys. Rev. Lett. \textbf{133}, 070403 (2024).

\bibitem{Berry-encircling}M. V. Berry, ``Optical polarization evolution near a nonHermitian degeneracy,'' J. Opt. A \textbf{13}, 115701 (2011).
\bibitem{Xu-encircling}H. Xu, D. Mason, Luyao Jiang, and J. G. E. Harris, ``Topological energy transfer in an optomechanical system with exceptional points,'' Nature (London) \textbf{537}, 80 (2016).
\bibitem{Li-encircling} A. Li,  J. Dong, J. Wang,  Z. Cheng,  J. S. Ho,  D. Zhang, et al., ``Hamiltonian hopping for efficient chiral mode switching in encircling exceptional points,'' Phys. Rev. Lett. \textbf{125}(18), 187403 (2020).
\bibitem{Feilhauer-encircling}J. Feilhauer, A. Schumer, J. Doppler, A. A. Mailybaev, J. B\"{o}hm, U. Kuhl,  N. Moiseyev, and S. Rotter, ``Encircling exceptional points as a non-Hermitian extension of rapid adiabatic passage,'' Phys. Rev. A \textbf{102}, 040201 (2020).
\bibitem{Ergoktas-encircling} M. S. Ergoktas, S. Soleymani, N. Kakenov, K. Wang, T. B. Smith, G. Bakan, S. Balci, A. Principi, K. S. Novoselov, S. K. Ozdemir, and C. Kocabas, ``Topological engineering of terahertz light using electrically tunable exceptional point singularities,'' Science \textbf{376}, 184 (2022).
\bibitem{Arkhipov-encircling1} I. I. Arkhipov, A. Miranowicz, F. Minganti, $\c{S}$. \"{O}zdemir, and F. Nori, ``Dynamically crossing diabolic points while encircling exceptional curves: A programmable symmetricasymmetric multimode switch,'' Nat. Commun. \textbf{14}, 2076 (2023).

\bibitem{Tang-encircling}Z. Tang, T. Chen, and X. Zhang, ``Highly efficient transfer of quantum state and robust generation of entanglement state around exceptional lines,'' Laser Photonics Rev. 2300794 (2023).
\bibitem{Hassan-encircling}A. U. Hassan, G. L. Galmiche, G. Harari and P. LiKamWa, ``Chiral state conversion without encircling an exceptional point,'' Phys. Rev. A \textbf{96}, 052129 (2017).
\bibitem{Hassan-encircling2}A. U. Hassan, B. Zhen, M. Solja\v{c}i\'{c}, M. Khajavikhan, and D. N. Christodoulides, ``Dynamically encircling exceptional points: Exact evolution and polarization state conversion,'' Phys. Rev. Lett. \textbf{118}, 093002 (2017).
\bibitem{Zhang-encircling} X. L. Zhang, T. Jiang, and C. T. Chan, ``Dynamically encircling an exceptional point in anti-parity-time symmetric systems: Asymmetric mode switching for symmetrybroken modes,'' Light Sci. Appl. \textbf{8}, 88 (2019).
\bibitem{Nasari-encircling}H. Nasari, G. L. Galmiche, H. E. L. Aviles, A. Schumer, A. U. Hassan, Q. Zhong, S. Rotter, P. LiKamWa, D. N. Christodoulides and M. Khajavikhan, ``Observation of chiral state transfer without encircling an exceptional point,'' Nature \textbf{605}, 256 (2022).

\bibitem{open1}F. Verstraete,  M. M. Wolf and  J. I. Cirac, ``Quantum computation and quantum-state engineering driven by dissipation,'' Nat. Phys. \textbf{5}(9), 633 (2009).
\bibitem{open2}Q. C. Wu, Y. H. Zhou, B. L. Ye, T. Liu and C. P. Yang, ``Nonadiabatic quantum state engineering by time-dependent decoherence-free subspaces in open quantum systems,'' New J. Phys. \textbf{23}, 113005 (2021).

\bibitem{Bellstate1}J. Zou, S. Zhang, and Y. Tserkovnyak, ``Bell-state generation for spin qubits via dissipative coupling,''  Phys. Rev. B \textbf{106}, 180406 (2022).
\bibitem{Bellstate2} N. Schine,  A. W. Young,  W. J. Eckner,  M. J. Martin and A. M. Kaufman,  ``Long-lived Bell states in an array of optical clock qubits,''  Nat. Phys. \textbf{18}, 1067-1073 (2022).

\bibitem{JC2}Y. H. Zhou, H. Z. Shen, X. Y. Zhang, and X. X. Yi, ``Zero eigenvalues of a photon blockade induced by a non-Hermitian Hamiltonian with a gain cavity,'' Phys. Rev. A \textbf{97}, 043819 (2018).
\bibitem{JC1}B. W. Shore and P. L. Knight,  ``The jaynes-cummings model,'' J.  Mod. Optic. \textbf{40}, 1195, (1993).
\bibitem{JC3}L. S. Bishop, E. Ginossar and S. M. Girvin, ``Response of the Strongly Driven Jaynes-Cummings Oscillator,'' Phys. Rev. Lett. \textbf{105}, 100505  (2010).
\bibitem{JC4}C. Liu, J. F. Huang, ``Quantum phase transition of the Jaynes-Cummings model,'' Sci. China-Phys. Mech. Astron. \textbf{67}(1), 240312 (2024).

\bibitem{Singleexcitation1}J. Li, R. Yu, and Y. Wu, ``Proposal for enhanced photon blockade in parity-time-symmetric coupled microcavities,'' Phys. Rev. A \textbf{92}, 053837 (2015).
\bibitem{experiment-realize4}L. Feng, Z. J. Wong, R.-M. Ma, Y. Wang, and X. Zhang, ``Single-mode laser by parity-time symmetry breaking,'' Science \textbf{346}, 972 (2014).
\bibitem{experiment-realize5}W. Chen, $\c{S}$. \"{O}zdemir, G. Zhao, J. Wiersig, and L. Yang, ``Exceptional points enhance sensing in an optical microcavity,'' Nature (London) \textbf{548}, 192 (2017).
\bibitem{experiment-realize6}I. I. Arkhipov, A. Miranowicz, F. Nori, $\c{S}$. K. \"{O}zdemir, and F. Minganti, ``Fully solvable finite simplex lattices with open boundaries in arbitrary dimensions,'' Phys. Rev. Res. \textbf{5}, 043092 (2023).


\end{thebibliography}
\end{document}